\documentclass[12pt]{article}

\usepackage{epsf}
\usepackage[dvips]{graphicx}

\begin{document}

\begin{center}
{\bfseries FURTHER EVIDENCE OF NARROW BARYONIC STRUCTURES WITH HADRONIC 
AS WELL AS LEPTONIC PROBES}

\vskip 5mm

B. Tatischeff$^{1}$ and E. Tomasi-Gustafsson$^{2}$

\vskip 5mm

{\small
{\it
$^{1)}$Institut de Physique Nucl\'eaire,CNRS/IN2P3, F--91406 Orsay Cedex,
France}\\
{\it
$^{2)}$DAPNIA/SPhN, CEA/Saclay, 91191 Gif-sur-Yvette Cedex, France}\\

$\dag$ {\it
E-mail: tati@ipno.in2p3.fr,\hspace*{.2cm}etomasi@cea.fr}}
\end{center}

\vskip 5mm

\begin{center}
\begin{minipage}{150mm}
\centerline{\bf Abstract}
Although extracted from several experiments using hadronic probes \cite{bor1},
narrow baryonic structures have been sometimes met with disbelief. New
signatures are presented, which appear from already published data, 
obtained with hadronic probes as well as with leptonic probes. The authors
of these results did not take into account the possibility to associate the discontinuities
of their spectra with the topic of narrow baryonic low mass structures. The stability of 
the observed narrow structure masses, represents a confirmation of their genuine existence.
\end{minipage}
\end{center}

\vspace*{-10mm}
\section{Narrow baryons produced by hadrons}
Using mainly the pp$\to$p$\pi^{+}$X and pp$\to$ppX reactions, studied at SPES3 
(Saturne), a spectrum of narrow baryonic structures was observed \cite{bor1}.
A high statistic missing mass spectrum of the
p($\alpha,\alpha$')X reaction (T$_{\alpha}$=4.2~GeV, $\theta$=0.8$^{0}$)
was obtained twelve years ago at SPES4 (Saturne) in order to
study the radial excitation of the nucleon in the P$_{11}$(1440 MeV) Roper
resonance \cite{mor}. A first large peak around $\omega\approx$~240~MeV was
associated with the projectile excitation, and a second large peak around
$\omega\approx$~510~MeV was associated with the target excitation. Above them
lie narrow peaks, defined by a large number of standard deviations (see
Fig.~1). Their masses (see table~1) agree fairly well with the masses of
narrow structures extracted from
pp$\to$p$\pi^{+}$ and pp$\to$ppX reactions studied at SPES3
(Saturne) \cite{bor1}. Fig.~2 shows the spectra of the same reaction at
$\theta$=2$^{0}$ \cite{mor1}. The empty circles, in both figures, which 
correspond to the scale, are the published number of events versus the
energy loss. The full circles and full squares show the same data in an
expanded scale \cite{bor2}.
Table 1 and Fig.~3 give the quantitative informations concerning the masses
extracted from the previous figures,
and the comparison with the masses previously extracted from SPES3 cross-sections
\cite{bor1}. Nearly all peaks are seen in both experiments. At
$\theta$=0.8$^{0}$, the incident beam enters the SPES4 spectrometer,
preventing a possible confirmation of the lower mass structure at
M=1004~MeV. Above M=1470~MeV, a lot of peaks are observed. The same situation
is observed in the SPES3 data \cite{bor3}, and all masses observed in both
reactions are about the same. The peak at M=1394~MeV
 observed in the SPES4 experiment, was not kinematically accessible in the SPES3 data,
since the mass range 1400$\le$M$\le$1470~MeV lie between two incident proton
energies.\\
\hspace*{4.mm} We observe a nice agreement between the masses obtained
using data from different physicists, studying different reactions with
different probes and different experimental equipements.\\ 
\hspace*{4.mm} This correlation is shown in Fig.~3, where the masses of the
structures observed at SPES4 are shown versus the ones at SPES3. The
straight lines correspond to the same masses, and all points are  
located along these lines. Peaks observed in one experiment only are shown as
empty circles.
Some other structures were observed in a few p(d,d')X spectra,
which masses correspond with a high accuracy to the SPES3 masses.  
\begin{figure}[ht]
\hspace*{-.5cm}
\centerline{
\includegraphics[width=190mm,height=56mm]{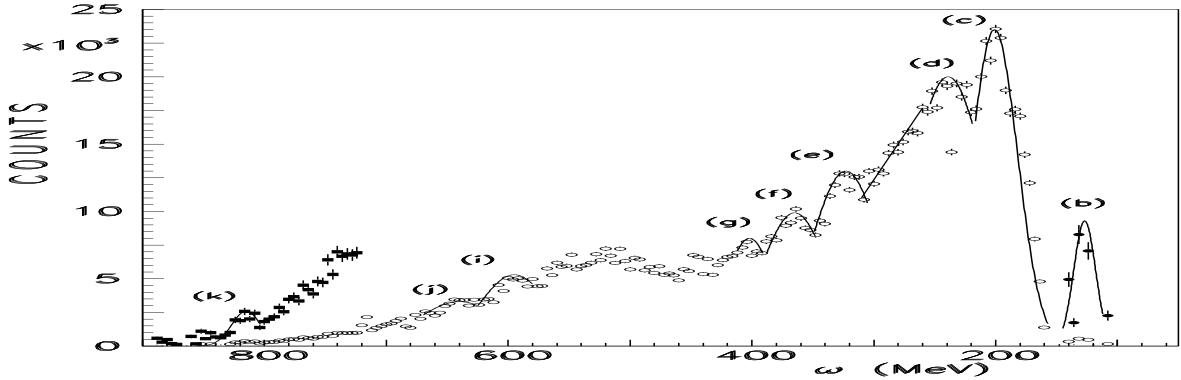}}
\begin{center}
\caption{Spectra of the p($\alpha,\alpha'$)X reaction studied at SPES4
(Saturne) with T$_{\alpha}$=4.2~GeV and $\theta$=0.8$^{0}$ [2].}
\end{center}
\end{figure}
\vspace*{-5.mm}
\begin{figure}[h]
\vspace*{-.5cm}
\hspace*{-.5cm}
\centerline{
\includegraphics[width=190mm,height=56mm]{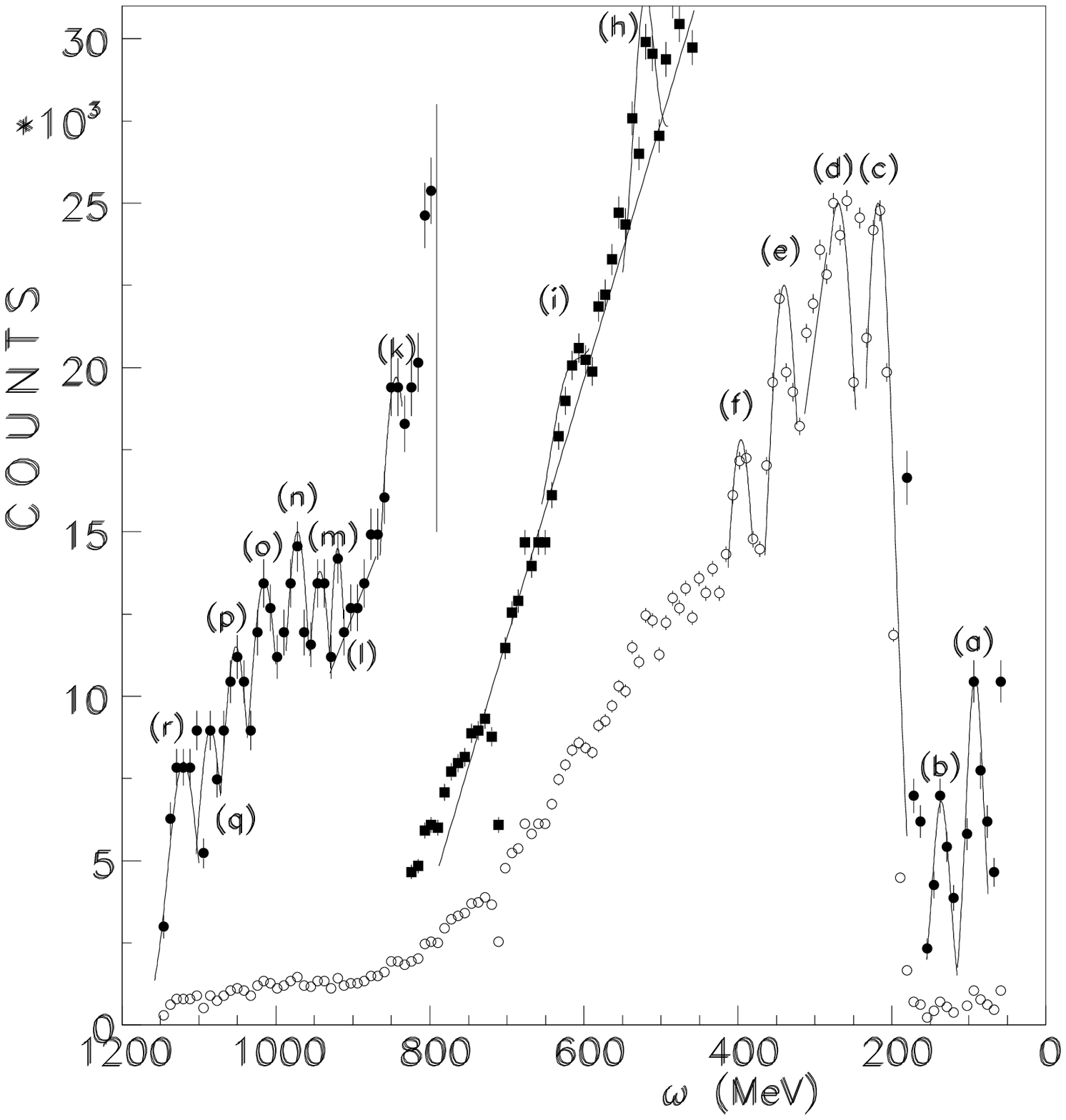}}
\begin{center}
\caption{Spectra of the p($\alpha,\alpha'$)X reaction studied at SPES4
(Saturne) with T$_{\alpha}$=4.2~GeV and $\theta$=2$^{0}$ [3].}
\end{center}
\end{figure}
\begin{figure}[h]
\vspace*{-5.mm}
\epsfysize=95mm
\centerline{
\includegraphics[width=190mm,height=135mm]{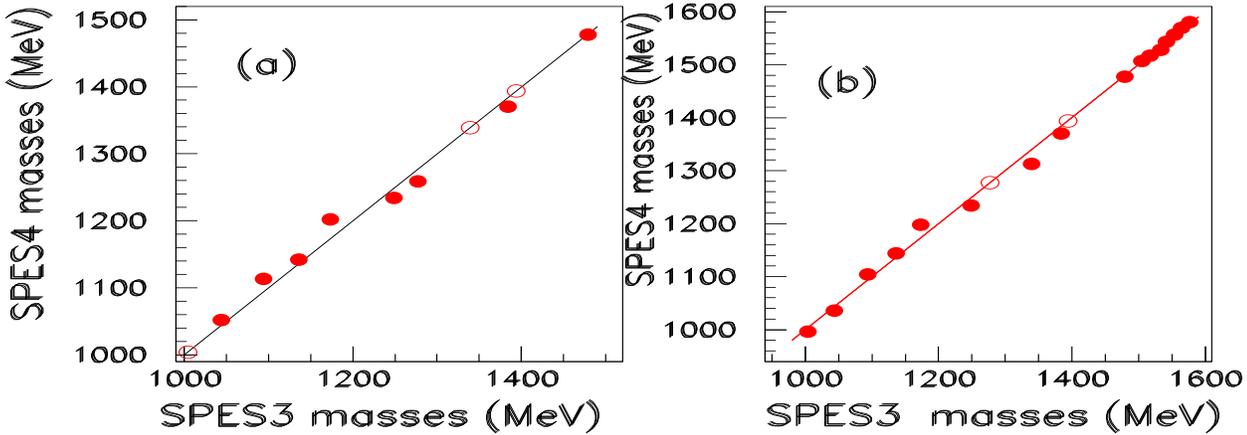}}
\vspace*{-60.mm}
\caption{Comparison between masses of narrow baryons extracted from
SPES3 and SPES4 data. Inserts (a) and (b) correspond respectively to
$\theta$=0.8$^{0}$ and $\theta$=2$^{0}$.}
\end{figure}
\begin{table}[h]
\caption{Masses (in MeV) of narrow exotic baryons, observed previously in
SPES3 data
and extracted from previous p($\alpha,\alpha$')X spectra measured at SPES4
[2] [3].}
\label{Table 1}
\vspace*{.3cm}

\hspace*{-.6cm}
\begin{tabular} [h]{c c c c c c c c c c c c}
\hline
SPES3 mass&1004&1044&1094&1136&1173&1249&1277&1339&1384&&1479\\
pic marker&(a)&(b)&(c)&(d)&(e)&(f)&(g)&(h)&(i)&(j)&(k)\\
SPES4 mass 0.8$^{0}$&&1052&1113&1142&1202&1234&1259&1370&1394&&1478\\
SPES4 mass 2$^{0}$&996&1036&1104&1144&1198&1234&1313&1370&&&1477\\
\hline
\hline
SPES3 mass&1505&1517&1533&1542&(1554)&1564&1577&&&&\\
pic marker&l&(m)&(n)&(o)&(p)&(q)&(r)&&&\\
SPES4 mass 2$^{0}$&1507&1517&1530&1543&1557&1569&1580&&&&\\
\hline
\end{tabular}
\end{table}
\section{Narrow baryons produced by leptons}
Recent precise attempts, fail to point out the narrow baryonic structures
observed with leptonic probes below pion threshold \cite{jiang} \cite{kohl}
\cite{zol}. However, at masses above pion threshold, as it was done before
with hadronic probes, several narrow structures can be extracted from
previous experiments performed to study other topics. A more detailed
paper which reviews similar results is in progress \cite{egle}. In order
to illustrate the previous comment, we show four different data in the next
figures.\\
\hspace*{4.mm}Among the numerous experiments of Compton scattering  on proton, 
some results from Saskatchewan \cite{hal} are shown in Fig.~4. A peak not discussed
by the authors, is easily extracted at M$\approx$1094~MeV, at the same mass
where it was seen before in the pp$\to$p$\pi^{+}$X reaction at SPES3.
Inserts (a) and (b) correspond to $\theta^{0}_{c.m.}$=90$^{0}$  and
141$^{0}$ respectively, with a peak at T$_{\gamma}$=166.5 (169)~MeV. 
The total c.m. energies are respectively $\sqrt{s}$=1092.2 and 1094.3~MeV,
the width of the peaks is $\sigma$=5.7 (5.2)~MeV and the number of standard deviations S.D.=3.3 (5.5).\\
\begin{figure}[h]
\vspace*{-2.cm}
\epsfysize=130mm
\centerline{
\includegraphics[width=190mm,height=110mm]{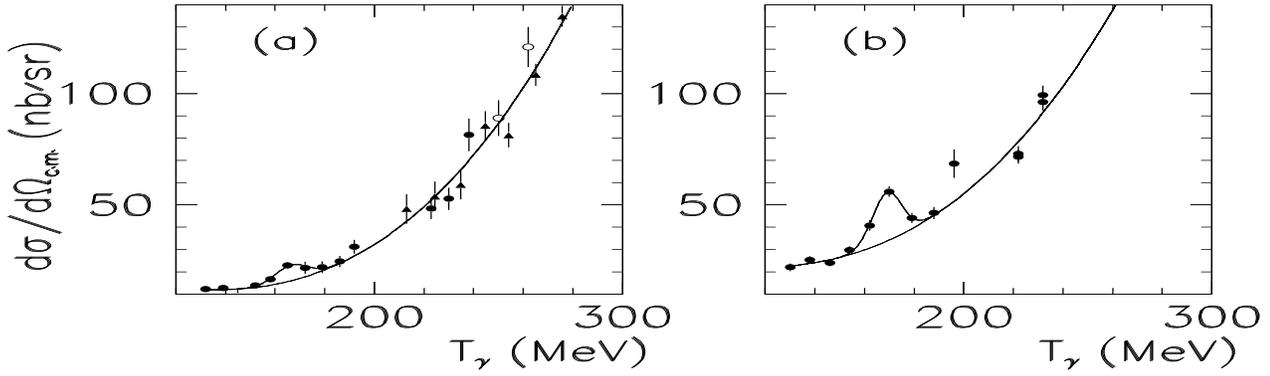}}
\vspace*{-5.cm}
\caption{Selection of Compton scattering from the proton data, measured at
Saskatchewan [10] (see [9]).}
\end{figure}
\begin{figure}[h]
\epsfysize=130mm
\centerline{
\includegraphics[width=190mm,height=110mm]{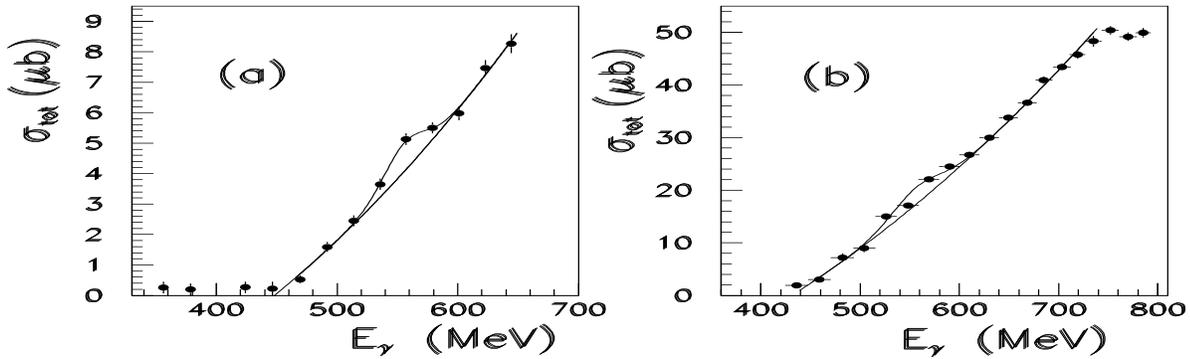}}
\vspace*{-5.cm}
\caption{Total cross-sections of two pion production measured at MAMI.
Insert (a) shows $\sigma_{tot}$ of $\gamma$p$\to\pi^{0}\pi^{0}$p [11];
insert (b) shows $\sigma_{tot}$ of $\gamma$n$\to\pi^{-}\pi^{0}$p [12].
The quantitative informations are given in [9].}
\end{figure}
\hspace*{4.mm}Fig.~5 illustrates total cross-sections of two pion
photoproduction measured at MAMI, namely $\gamma$p$\to\pi^{0}\pi^{0}$p
\cite{hart} and $\gamma$n$\to\pi^{-}\pi^{0}$p \cite{zab}. Both peaks
correspond to $\surd$s=1387~MeV, close to M=1384~MeV (mass of a narrow
structure already seen \cite{bor2}).\\
\hspace*{4.mm}Fig.~6 illustrates total cross-sections of one pion
photoproduction measured at INS (Tokyo) \cite{fujii}. Inserts (a) and (b)
correspond respectively to the $\gamma$p$\to\pi^{+}$n and
$\gamma$n$\to\pi^{-}$p reactions. Insert (a) shows a peak at M=1389~MeV,
(M=1384~MeV observed at SPES3). Insert (b) shows peaks at M=1171~MeV (1173),
M=1252~MeV (1249), and M=1387~MeV (1384).\\
\begin{figure}[h]
\vspace*{-8.mm}
\epsfysize=100mm
\centerline{
\includegraphics[width=160mm,height=98mm]{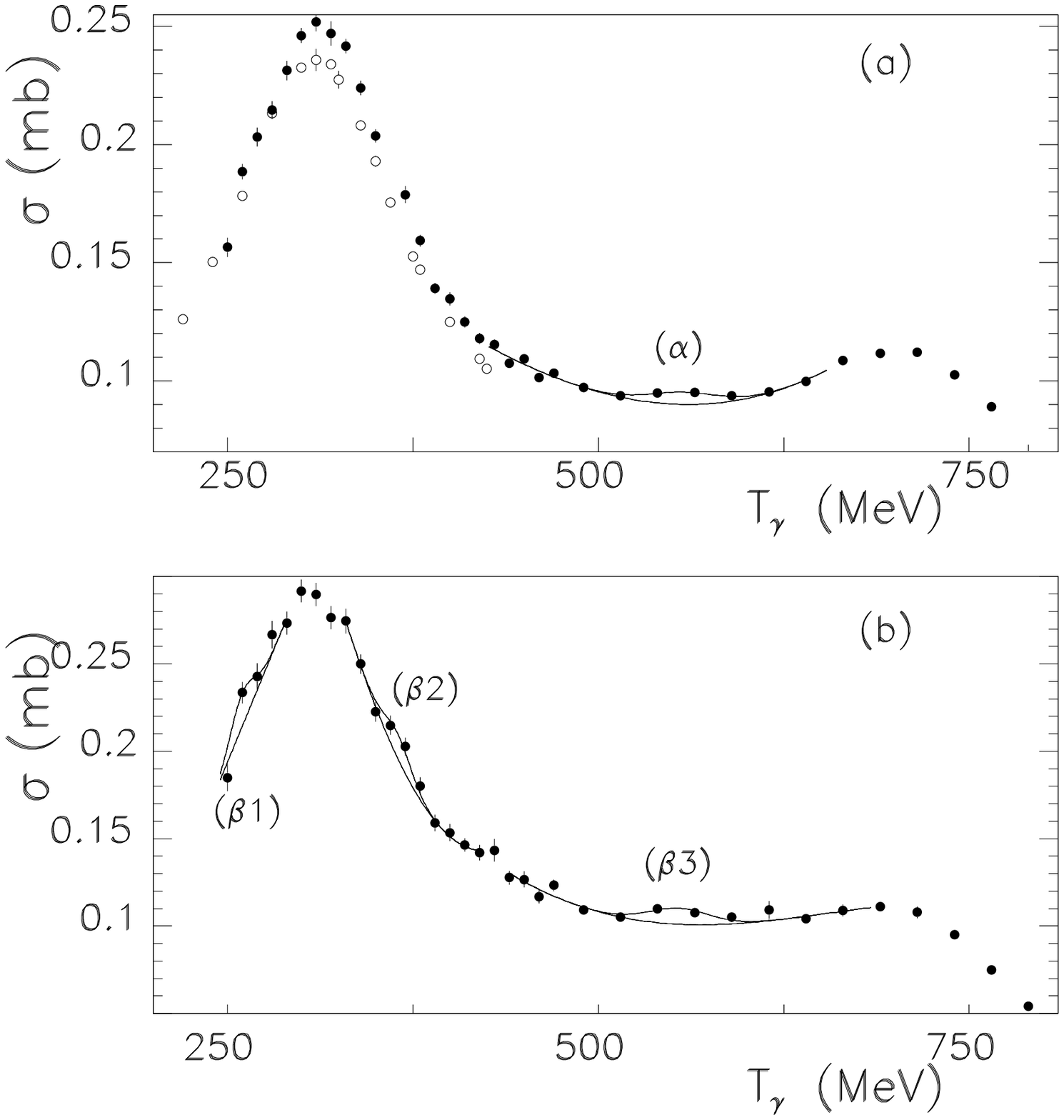}}
\caption{Total cross-sections of the $\gamma$p$\to\pi^{+}$n and
$\gamma$n$\to\pi^{-}$p reactions measured at the INS (Tokyo).
The quantitative informations are given in [9].}
\end{figure}
\begin{figure}[h]
\vspace*{-18.mm}
\epsfysize=100mm
\centerline{
\includegraphics[width=160mm,height=90mm]{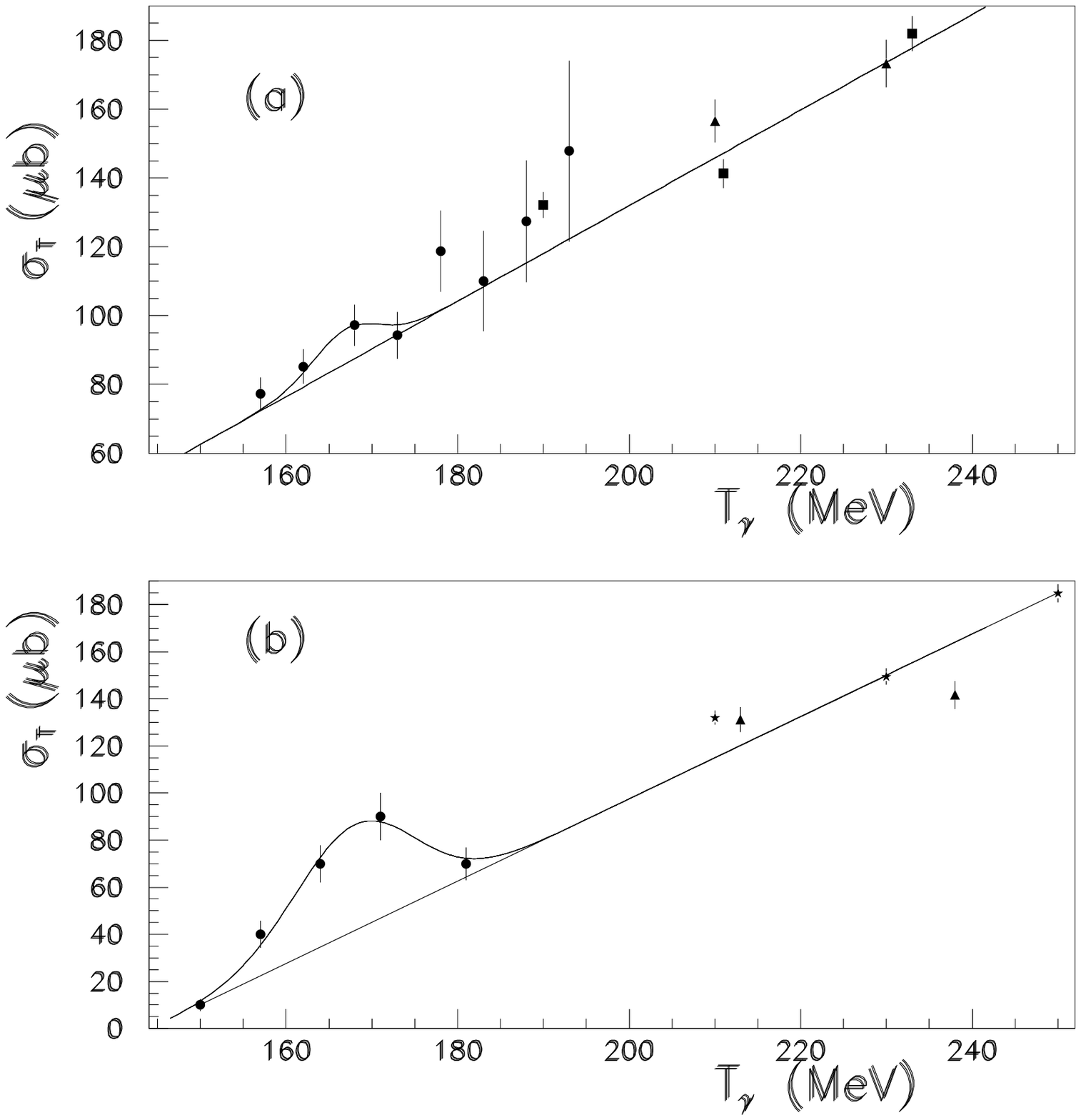}}
\caption{Total cross-sections of the $\gamma$n$\to\pi^{-}$p (insert (a)) 
and $\gamma$d$\to\pi^{-}$pp (insert (b)) reactions measured at various laboratories. 
The quantitative informations are given in [9].}
\end{figure}
\hspace*{4.mm}Fig.~7 illustrates total cross-sections of one pion
photoproduction on $^{1}$H and $^{2}$H targets. The
$\gamma$n$\to\pi^{-}$p total cross-section data are from different 
laboratories \cite{gold}, when the $\gamma$d$\to\pi^{-}$pp total 
cross-sections are quoted from the CERN-Hera Compilation \cite{adam}.
Both spectra exhibit a peak at M=1094~MeV and M=1095~MeV, to compare to
M=1094~MeV (a SPES3 peak mass).\\
\hspace*{4.mm}Other examples can be given \cite{egle}, sometimes the peaks
are smaller than those
which are illustrated in this paper; they always show an astonishing
correspondance with the masses previously 
extracted with hadronic probes.\\
\hspace*{4.mm}Fig.~8 shows a comparison of masses observed with leptons,
versus masses
observed with hadrons. Here the solid symbols give this comparison and the
empty circles show the peak masses not observed with leptonic probes. We
observe less structures extracted from reactions with leptonic probes, but 
when a peak was seen, its mass reproduces well a peak mass from SPES3
experiments.
\begin{figure}[h]
\vspace*{-3.5cm}
\epsfysize=135mm
\centerline{
\includegraphics[width=160mm,height=125mm]{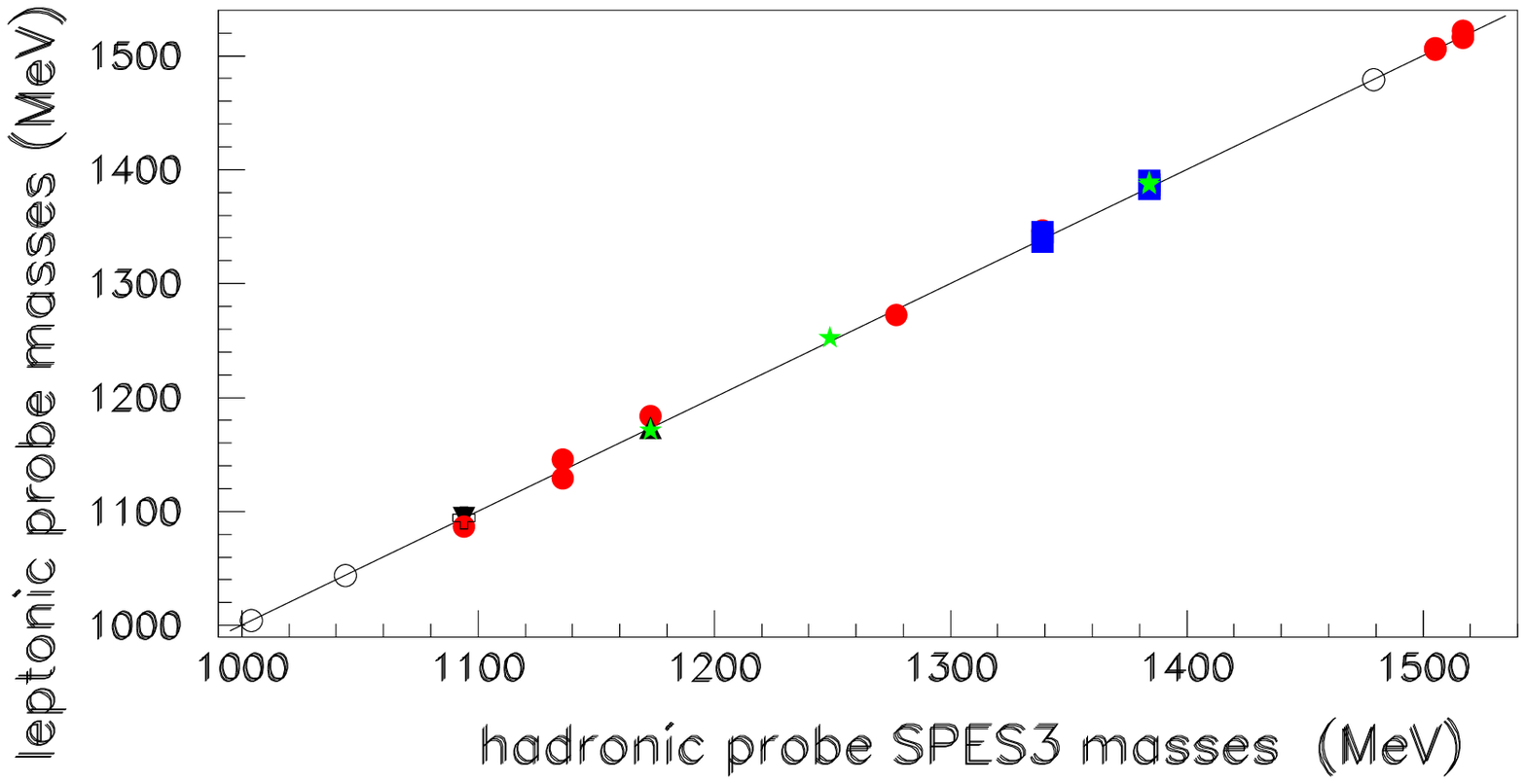}}
\vspace*{-4.5cm}
\caption{Comparison of masses of narrow baryonic structures observed with
leptonic probes, versus masses of narrow baryonic structures observed with
hadronic probes. The empty circles show data not seen with incident leptons.}
\end{figure}
\section{Conclusion}
New dedicated precise experiments must be done to confirm the masses of the
narrow baryons observed in many previous experiments, mainly performed at
Saturne on the SPES3 and SPES4 beam lines. However, the stability
of the structures already extracted from data obtained to study other topics,
is noteworthy. This comment concerns as well data obtained with hadronic
probes as
data obtained with leptonic probes, although these last are less numerous.
It is worthwhile to mention that the few baryons with mass lower than the
pion threshold, were not seen in recent experiments using leptonic probes. 
Then it is possible to speculate that the excitation of these baryons is favoured
through dibaryonic states.

\end{document}